\documentclass[%
reprint,
superscriptaddress,
showpacs,
amsmath,amssymb,
prc,
floatfix, ]%
{revtex4-1}

\usepackage{color}

\usepackage{graphicx}% Include figure files
\usepackage{dcolumn}% Align table columns on decimal point
\usepackage{bm}% bold math
\usepackage[dvipdfm,bookmarks=true,colorlinks,%
            citecolor=blue,linkcolor=blue,hypertex, %
            breaklinks=true]{hyperref}

\begin{document}

%\begin{CJK*}{GBK}{}

%\preprint{NP-ITP/CAS-201002}

\title{Particle-number conserving analysis of rotational bands
       in $^{247, 249}$Cm and $^{249}$Cf}

\author{Zhen-Hua Zhang}% (͖́Ȼ)}%
 \affiliation{Key Laboratory of Frontiers in Theoretical Physics,
              Institute of Theoretical Physics, Chinese Academy of Sciences,
              Beijing 100190, China}
\author{Jin-Yan Zeng}% (Ôø½÷ÑÔ)}%
 \affiliation{School of Physics, Peking University,
              Beijing 100871, China}
\author{En-Guang Zhao}% (ÕÔ¶÷¹ã)}%
% \email{egzhao@itp.ac.cn}
 \affiliation{Key Laboratory of Frontiers in Theoretical Physics,
              Institute of Theoretical Physics, Chinese Academy of Sciences,
              Beijing 100190, China}
 \affiliation{Center of Theoretical Nuclear Physics, National Laboratory
              of Heavy Ion Accelerator, Lanzhou 730000, China}
 \affiliation{School of Physics, Peking University,
              Beijing 100871, China}
\author{Shan-Gui Zhou}% (ÖÜÉƹó)}
 \email{sgzhou@itp.ac.cn}
 \homepage{http://www.itp.ac.cn/~sgzhou}
 \affiliation{Key Laboratory of Frontiers in Theoretical Physics,
              Institute of Theoretical Physics, Chinese Academy of Sciences,
              Beijing 100190, China}
 \affiliation{Center of Theoretical Nuclear Physics, National Laboratory
              of Heavy Ion Accelerator, Lanzhou 730000, China}

\date{\today}

\begin{abstract}
The recently observed high-spin rotational bands in odd-$A$ nuclei
$^{247, 249}$Cm and $^{249}$Cf [Tandel \textit{et al.}, Phys. Rev. C
82 (2010) 041301R] are investigated by using the cranked shell model
(CSM) with the pairing correlations treated by a particle-number
conserving (PNC) method in which the blocking effects are taken into
account exactly. The experimental moments of inertia and alignments
and their variations with the rotational frequency $\omega$ are
reproduced very well by the PNC-CSM calculations. By examining the
$\omega$-dependence of the occupation probability of each cranked
Nilsson orbital near the Fermi surface and the contributions of
valence orbitals to the angular momentum alignment in each major
shell, the level crossing and upbending mechanism in each nucleus is
understood clearly.
\end{abstract}

%21.10.-k 	Properties of nuclei; nuclear energy levels
%21.60.-n 	Nuclear structure models and methods
%21.60.Cs 	Shell model
%23.20.Lv 	¦Ã transitions and level energies
%27.90.+b 	A ¡Ý 220


\maketitle

%\section{Introduction}

The rotational spectra of the heaviest nuclei can reveal
detailed information on the single particle configurations, the
shell structure, the stability against rotation, \textit{etc.},
thus providing a benchmark for nuclear models. In recent years,
the in-beam spectroscopy of nuclei with $Z \approx 100$ has
been one of the most important frontiers on nuclear structure
physics~\cite{Herzberg2008}. Besides even-even
nuclei~\cite{Reiter1999,Herzberg2001,Bastin2006}, experimental
efforts have been also focused on the study of high-spin states
of odd-$A$ nuclei, such as $^{253}$No~\cite{Reiter2005} and
$^{251}$Md~\cite{Chatillon2007}. Quite recently, the rotational
bands of odd-$A$ $^{247,249}$Cm and $^{249}$Cf were observed up
to very high spins ($\approx 28 \hbar$) and appropriate single particle
configurations have been assigned to these bands
%unambiguously
~\cite{Tandel2010}. It is worthwhile to mention
that the neutron $\nu1/2^+$[620] band in $^{249}$Cm
% the rotational band in $^{249}$Cm is built on the
%$\nu1/2^+$[620] Nilsson orbital which
is the highest-lying
neutron configuration investigated up to very high spins.
Although the cranking Woods-Saxon calculations reproduced well some of the
observed properties, this experiment, together with some
previous ones, still challenge nuclear structure models,
\textit{e.g.}, the absence of the alignment of $j_{15/2}$
neutrons in several nuclei in this mass region needs a
consistent explanation~\cite{Tandel2010}.

In this paper, %we use
the cranked shell model (CSM) with %the
pairing correlations treated by a particle-number conserving
(PNC) method~\cite{Zeng1983} is used to investigate the rotational
bands in $^{247,249}$Cm and $^{249}$Cf observed in
Ref.~\cite{Tandel2010}. In contrary to the conventional BCS
approach, in the PNC method, the particle-number is conserved
and the Pauli blocking effects are taken into account exactly.
The PNC-CSM treatment has been used to describe successfully the normally
deformed and superdeformed high spin rotational bands of nuclei
with $A \approx$ 160, 190, and
250~\cite{Zeng1994,Zeng1994a,Liu2002a,He2005,Zhang2009,He2009a}.

%\section{A brief introduction to PNC}

The details of the PNC-CSM treatment can be found in
Refs.~\cite{Zeng1994,Zeng1994a}. For convenience, here we briefly give the related
formalism. The CSM Hamiltonian of an axially symmetric nucleus
in the rotating frame is
\begin{eqnarray}
 H_\mathrm{CSM}
 & = &
 H_0 + H_\mathrm{P}
 = H_{\rm Nil}-\omega J_x + H_\mathrm{P}
 \ ,
 \label{eq:H_CSM}
\end{eqnarray}
where $H_{\rm Nil}$ is the Nilsson Hamiltonian, $-\omega J_x$
is the Coriolis interaction with the cranking frequency
$\omega$ about the $x$ axis (perpendicular to the nuclear
symmetry $z$ axis). $H_{\rm P} = H_{\rm P}(0) + H_{\rm P}(2)$
is the pairing interaction,
\begin{eqnarray}
 H_{\rm P}(0)
 & = &
  -G_{0} \sum_{\xi\eta} a^+_{\xi} a^+_{\bar{\xi}}
                        a_{\bar{\eta}} a_{\eta}
  \ ,
 \\
 H_{\rm P}(2)
 & = &
  -G_{2} \sum_{\xi\eta} q_{2}(\xi)q_{2}(\eta)
                        a^+_{\xi} a^+_{\bar{\xi}}
                        a_{\bar{\eta}} a_{\eta}
  \ ,
\end{eqnarray}
where $\bar{\xi}$ ($\bar{\eta}$) labels the time-reversed state of a
Nilsson state $\xi$ ($\eta$), $q_{2}(\xi) = \sqrt{{16\pi}/{5}}
\langle \xi |r^{2}Y_{20} | \xi \rangle$ is the diagonal element of
the stretched quadrupole operator, and $G_0$ and $G_2$ are the
effective strengths of monopole and quadrupole pairing interactions,
respectively.

Instead of the usual single-particle level truncation in common
shell-model calculations, a cranked many-particle configuration
(CMPC) truncation (Fock space truncation) is adopted which is crucial
to make the PNC calculations for low-lying excited states both
workable and sufficiently accurate~\cite{Molique1997}. An eigenstate
of $H_\mathrm{CSM}$ can be written as
\begin{equation}
 |\Psi\rangle = \sum_{i} C_i \left| i \right\rangle
 \ ,
 \qquad (C_i \; \textrm{real}),
\end{equation}
where $| i \rangle$ is a CMPC (an eigenstate of the one-body operator
$H_0$). By diagonalizing the $H_\mathrm{CSM}$ in a sufficiently
large CMPC space, sufficiently accurate solutions for low-lying excited eigenstates of
$H_\mathrm{CSM}$ are obtained.

The kinematic moment of inertia (MOI) for the state $| \Psi
\rangle$ is
\begin{eqnarray}
 J^{(1)}
 & = &
  \frac{1}{\omega} \left\langle \Psi | J_x | \Psi \right\rangle
 \nonumber \\
 & = &
  \frac{1}{\omega} \left( \sum_i C_i^2 \left\langle i | J_x | i \right\rangle
                        +2\sum_{i<j} C_i C_j \left\langle i | J_x | j \right\rangle
                   \right)
 \ .
\end{eqnarray}
Considering $J_x$ to be a one-body operator, the matrix element
$\langle i | J_x | j \rangle$ for $i\neq j$ is nonzero only when
$|i\rangle$ and $|j\rangle$ differ by one particle
occupation~\cite{Zeng1994, Zeng1994a}. After a certain permutation of
creation operators, $|i\rangle$ and $|j\rangle$ can be recast into $|
i \rangle = (-1)^{M_{i\mu}} | \mu \cdots \rangle$ and $| j \rangle =
(-1)^{M_{j\nu}} | \nu \cdots \rangle$ where the ellipsis ``$\cdots$''
stands for the same particle occupation and
$(-1)^{M_{i\mu(\nu)}}=\pm1$ according to whether the permutation is
even or odd. Therefore, the angular momentum alignment of
$|\Psi\rangle$ can be expressed as
\begin{equation}
 \langle \Psi | J_x | \Psi \rangle = \sum_{\mu} j_x(\mu) + \sum_{\mu<\nu} j_x(\mu\nu)
 \ .
 \label{eq:jx}
\end{equation}
The diagonal contribution $j_x(\mu)=\langle \mu | j_{x} | \mu \rangle
n_{\mu}$ where $n_{\mu} = \sum_{i} |C_{i}|^{2} P_{i\mu}$ is the
occupation probability of the cranked Nilsson orbital $|\mu\rangle$
and $P_{i\mu}=1$ (0) if $|\mu\rangle$ is occupied (empty). The
off-diagonal (interference) contribution $j_x(\mu\nu)=2 \langle \mu |
j_{x} | \nu \rangle
  \sum_{i<j} (-1)^{M_{i\mu}+M_{j\nu}} C_{i} C_{j}$.

%\section{Results and discussions}

\begin{table}[!h]
\caption{\label{tab:ku}Nilsson
parameters $\kappa$ and $\mu$ proposed for nuclei with $A \approx$
250~\protect\cite{Zhang2010f}.}
\begin{center}
\def\temptablewidth{8.6cm}
% {\rule{\temptablewidth}{1pt}}
\begin{tabular*}%{cccc|cccc}%
{\temptablewidth}{@{\extracolsep{\fill}}cccc|cccc}
\hline\hline
 $N$ & $l$     & $\kappa_p$ & $\mu_p$ &
 $N$ & $l$     & $\kappa_n$ & $\mu_n$ \\
\hline
  4  & 0,2,4   & 0.0670     & 0.654   &
     &         &            &         \\
  5  & 1       & 0.0250     & 0.710   &
  6  & 0       & 0.1600     & 0.320   \\
     & 3       & 0.0570     & 0.800   &
     & 2       & 0.0640     & 0.200   \\
     & 5       & 0.0570     & 0.710   &
     & 4,6     & 0.0680     & 0.260   \\
  6  & 0,2,4,6 & 0.0570     & 0.654   &
  7  & 1,3,5,7 & 0.0634     & 0.318   \\
\hline\hline
\end{tabular*}
%{\rule{\temptablewidth}{1pt}}
\end{center}
\end{table}

The Nilsson parameters ($\kappa, \mu$)
systematics~\cite{Nilsson1969,Bengtsson1985a} can not reproduce well
the order of single particle levels for the very heavy nuclei with $A
\approx$ 250 (see, \textit{e.g.}, Ref.~\cite{He2009a}). Recently we
have proposed a new set of ($\kappa, \mu$) which is given in
Table~\ref{tab:ku} and deformation parameters for nuclei with $A
\approx$ 250 by fitting the observed single particle spectra of all
known odd-$A$ nuclei in this mass region~\cite{Zhang2010f}. Note that
the readjustment of Nilsson parameters is also necessary in some
other regions of the nuclear chart~\cite{Seo1986,Zhang1998}. The
deformation parameters ($\varepsilon_2, \varepsilon_4$) are (0.242,
0.002) for $^{247}$Cm and (0.248, 0.008) for $^{249}$Cm and
$^{249}$Cf. There are no experimental values of the deformation
parameters for these nuclei. The values used in various calculations
or predicted by different models are different. What we adopt here
are larger than those used in Ref.~\cite{Tandel2010} or predicted in
Ref.~\cite{Moller1995}, and smaller than the interpolated values from
Ref.~\cite{Al-Khudair2009} where, \textit{e.g.}, the $\epsilon_2$
values for $^{248}$Cf and $^{250}$Cf are 0.260 and 0.265
respectively.

The effective pairing strengths, in principle, can be determined by
the odd-even differences in binding energies, and are connected with
the dimension of the truncated CMPC space. The CMPC space for the
very heavy nuclei is constructed in the proton $N=4, 5, 6$ shells and
the neutron $N=6, 7$ shells. The dimensions of the CMPC space are
about 1000 for both protons and neutrons in our calculation. The
corresponding effective monopole and quadrupole pairing strengths are
$G_{0p}=0.50$~MeV and $G_{2p}=0.04$~MeV for protons,
$G_{0n}=0.30$~MeV and $G_{2n}=0.02$~MeV for neutrons. The stability
of the PNC calculation results against the change of the dimension of
the CMPC space has been investigated in Refs.~\cite{Zeng1994,
Liu2002a, Molique1997}. In the present calculations, almost all the
CMPC's with weight $>0.1\%$ are taken into account, so the solutions
to the low-lying excited states are accurate enough. A larger CMPC
space with renormalized pairing strengths gives essentially the same
results.

\begin{figure*}%[bthp]
\includegraphics[scale=0.6]{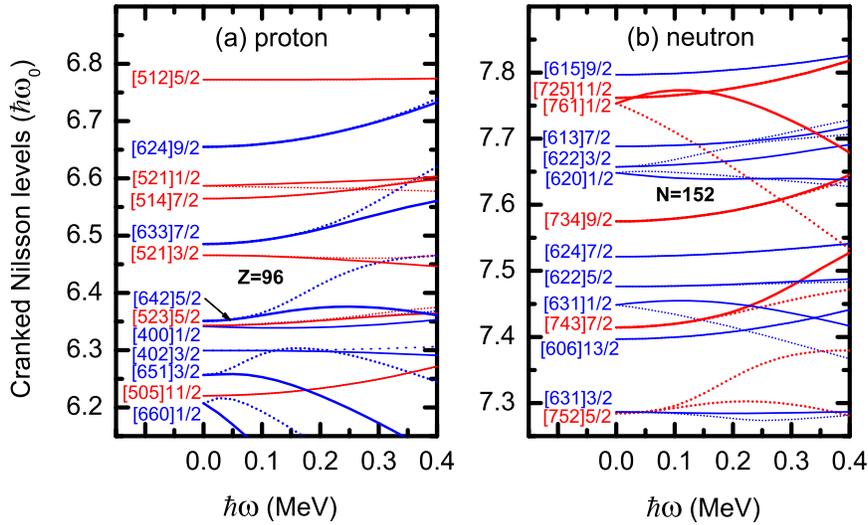}
\caption{\label{fig:Nilsson}(Color online) The cranked Nilsson
levels near the Fermi surface of $^{247}$Cm for protons (a) and
for neutrons (b). The positive (negative) parity levels are
denoted by blue (red) lines. The signature $\alpha=+1/2$
($\alpha=-1/2$) levels are denoted by solid (dotted) lines. }
\end{figure*}

Figure~\ref{fig:Nilsson} shows the calculated cranked Nilsson levels
near the Fermi surface of $^{247}$Cm. The positive (negative) parity
levels are denoted by blue (red) lines. The signature $\alpha=+1/2$
($\alpha=-1/2$) levels are denoted by solid (dotted) lines. For both
protons and  neutrons, the sequence of single-particle levels near
the Fermi surface is the same as the experimental data taken from
$^{247}$Cm and $^{247}$Bk~\cite{Herzberg2008} with the only exception
of the $\nu5/2^+[622]$ orbital. Many theoretical models predict that
the first excited state in $N = 151$ isotones should be
$\nu7/2^+[624]$ (see, \textit{e.g.}, Ref.~\cite{Parkhomenko2005}).
This is not consistent with experimental results, \textit{i.e.}, the
first excited state in $N=151$ isotones is $\nu5/2^+[622]$. The low
excitation energy of the $\nu5/2^+[622]$ state in the $N = 151$
isotones have been interpreted as a consequence of the presence of a
low-lying $K^\pi = 2^-$ octupole phonon state~\cite{Yates1975}.
Figure~\ref{fig:Nilsson} shows that there exist a proton gap at
$Z=96$ and  a neutron gap at $N=152$, which is consistent with the
experiment and the calculation by using a Woods-Saxon
potential~\cite{Chasman1977,*Chasman1978}. The cranked Nilsson levels
of $^{249}$Cm and $^{249}$Cf are quite similar to that of $^{247}$Cm
and not shown here.

\begin{figure*}%[ph]
\includegraphics[scale=0.80]{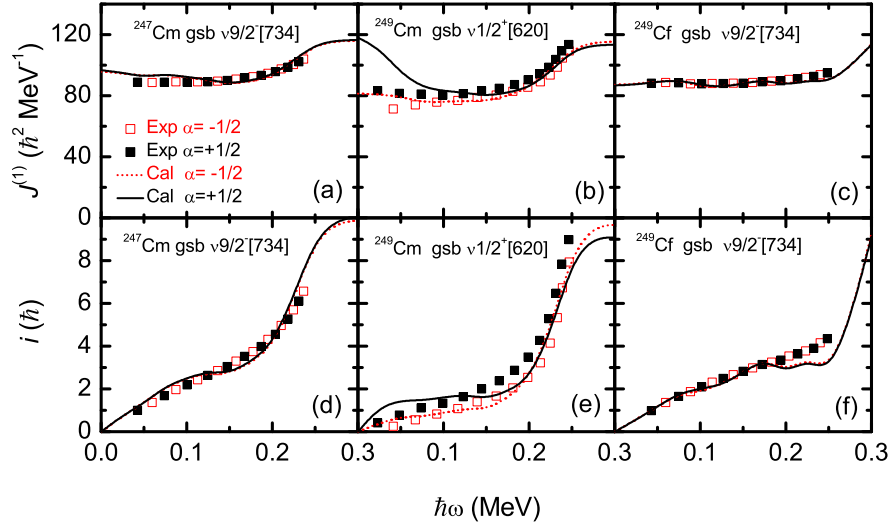}
\caption{\label{fig:MOI}(Color online) The experimental and
calculated MOIs $J^{(1)}$ and alignments (or called alignments
difference) of the ground state bands (gsb's) in $^{247, 249}$Cm and
$^{249}$Cf. The alignments difference $i$ is defined as $i= \langle
J_x \rangle -\omega J_0 -\omega ^ 3 J_1$ and the Harris parameters
$J_0 = 65\ \hbar^2$MeV$^{-1}$ and $J_1 = 200\ \hbar^4$MeV$^{-3}$ are
taken from Ref.~\protect\cite{Tandel2010}. The experimental MOIs and
alignments difference are denoted by solid squares (signature
$\alpha=+1/2)$ and open squares (signature $\alpha=-1/2)$,
respectively. The calculated MOIs and alignments difference are
denoted by solid lines (signature $\alpha=+1/2)$ and dotted lines
(signature $\alpha=-1/2)$, respectively. }
\end{figure*}

Figure~\ref{fig:MOI} shows the experimental and calculated MOIs and
alignments (see the caption of Fig.~\ref{fig:MOI}) of the ground
state bands (gsb's) in $^{247, 249}$Cm and $^{249}$Cf. The
experimental MOIs and alignments are denoted by solid squares
(signature $\alpha=+1/2$) and open squares (signature $\alpha=-1/2$),
respectively. The calculated MOIs and alignments are denoted by solid
lines (signature $\alpha=+1/2$) and dotted lines (signature
$\alpha=-1/2$), respectively. The experimental MOIs and alignments of
all these three 1-quasiparticle bands are well reproduced by the
PNC-CSM calculations, which in turn strongly support the
configuration assignments for these high-spin rotational bands
adopted in Ref.~\cite{Tandel2010}. Moreover, the signature splitting
in the $\nu1/2^+[620]$ band is also well reproduced by our
calculation, which is understandable from the behavior of the cranked
Nilsson orbital $\nu1/2^+[620]$ in Fig.~\ref{fig:Nilsson}. The
upbending frequency $\hbar\omega_c\sim 0.25$ for the gsb in
$^{249}$Cf is a little larger than that of the Cm isotopes
($\hbar\omega_c\sim$ 0.20~MeV). These results agree well with the
experiment and the cranking Woods-Saxon
calculations~\cite{Tandel2010}.

\begin{figure}%[h]
\includegraphics[scale=0.51]{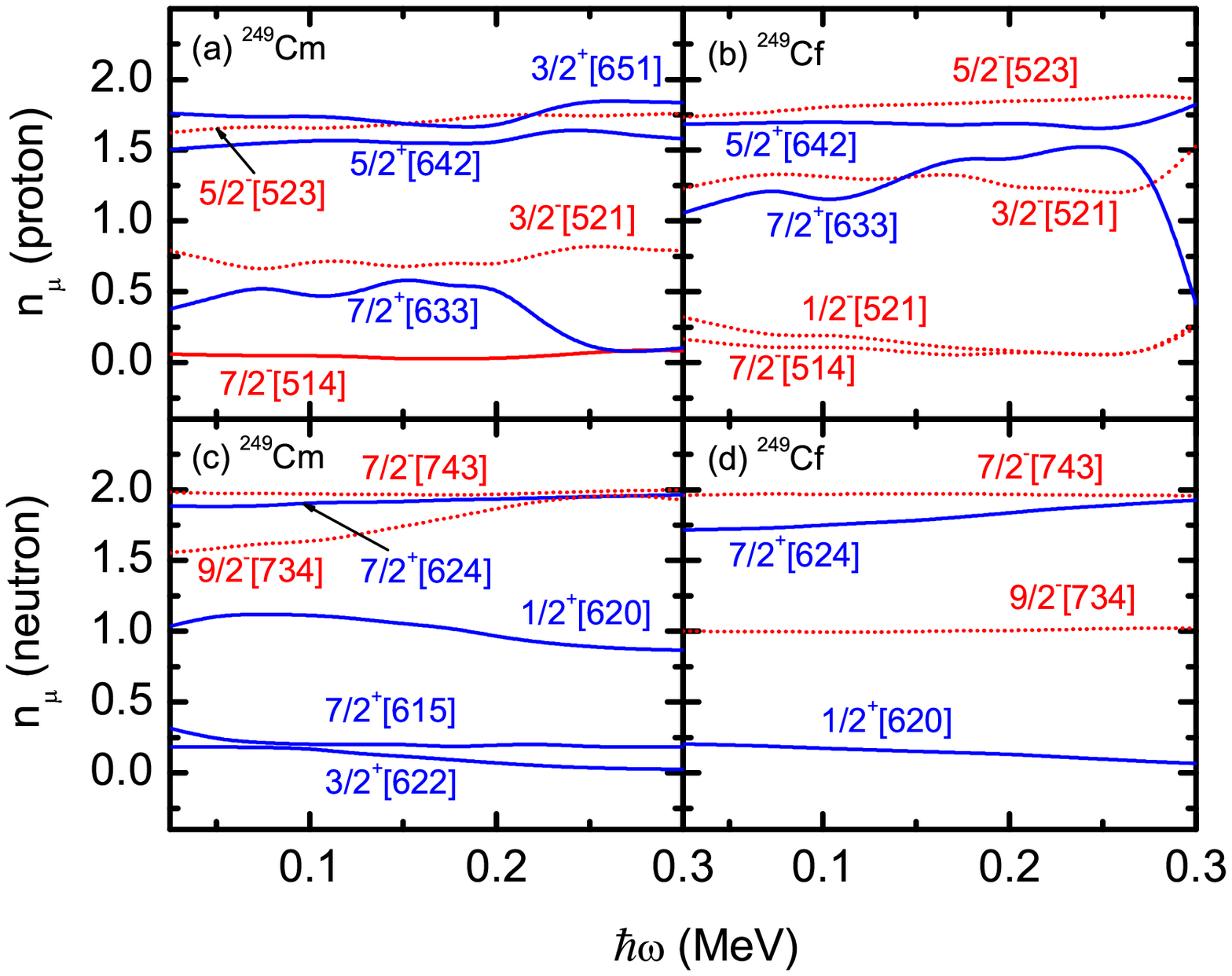}
\caption{\label{fig:Occupation}
(Color online) Occupation probability $n_\mu$ of each orbital
$\mu$ (including both $\alpha=\pm1/2$) near the Fermi surface
for the gsb's in $^{249}$Cm and $^{249}$Cf. The top and bottom
rows are for protons and neutrons respectively. The positive
(negative) parity levels are denoted by blue solid (red dotted)
lines. The Nilsson levels far above the Fermi surface
($n_{\mu}\sim0$) and far below ($n_{\mu}\sim2$) are not shown.
For the $\nu9/2^-[734]$ band in ${}^{247}$Cm, $n_\mu$ of proton
(neutron) orbitals are not shown because they are nearly the
same as those of ${}^{249}$Cm (${}^{249}$Cf).}
\end{figure}

One of the advantages of the PNC method is that the total particle
number $N = \sum_{\mu}n_\mu$ is exactly conserved, whereas the
occupation probability $n_\mu$ for each orbital varies with
rotational frequency $\hbar\omega$. By examining the
$\omega$-dependence of the orbitals close to the Fermi surface, one
can learn more about how the Nilsson levels evolve with rotation and
get some insights on the upbending mechanism.
Figure~\ref{fig:Occupation} shows the occupation probability $n_\mu$
of each orbital $\mu$ near the Fermi surface for the gsb's in
$^{249}$Cm and $^{249}$Cf. The top and bottom rows are for the
protons and neutrons respectively. The positive (negative) parity
levels are denoted by blue solid (red dotted) lines. We can see from
Fig.~\ref{fig:Occupation}(a) that the occupation probability of
$\pi7/2^+[633]$ ($i_{13/2}$) drops down gradually from 0.5 to nearly
zero with the cranking frequency $\hbar\omega$ increasing from about
$0.20$~MeV to $0.30$~MeV, while the occupation probabilities of some
other orbitals slightly increase. This can be understood from the
cranked Nilsson levels shown in Fig.~\ref{fig:Nilsson}(a). The
$\pi7/2^+[633]$ is slightly above the Fermi surface at
$\hbar\omega=0$. Due to the pairing correlations, this orbital is
partly occupied. With increasing $\hbar\omega$, this orbital leave
farther above the Fermi surface. So after the band-crossing
frequency, the occupation probability of this orbital becomes smaller
with increasing $\hbar\omega$. Meanwhile, the occupation
probabilities of those orbitals which approach near to the Fermi
surface become larger with increasing  $\hbar\omega$. This phenomenon
is even more clear in Fig.~\ref{fig:Occupation}(b), but the
band-crossing occurs at $\hbar\omega_c\sim0.25$~MeV, a little larger
than that of ${}^{249}$Cm. So the band-crossings in both cases are
mainly caused by the $\pi i_{13/2}$ orbitals. In
Fig.~\ref{fig:Occupation}(c), with increasing $\hbar\omega$ the
occupation probability of $\nu 1/2^+[620]$ decreases slowly and that
of the high-$\Omega$ (deformation aligned) $\nu 9/2^-[734]$ orbital
($j_{15/2}$) increases slowly. Thus only a small contribution is
expected from neutrons to the upbending for the gsb in $^{249}$Cm. In
Fig.~\ref{fig:Occupation}(d), the neutron orbital $\nu 9/2^-[734]$ of
$j_{15/2}$ parentage is totally blocked by an odd neutron, so it has
no contribution to the upbending for the gsb in ${}^{249}$Cf.

\begin{figure}%[h]
\includegraphics[scale=0.51]{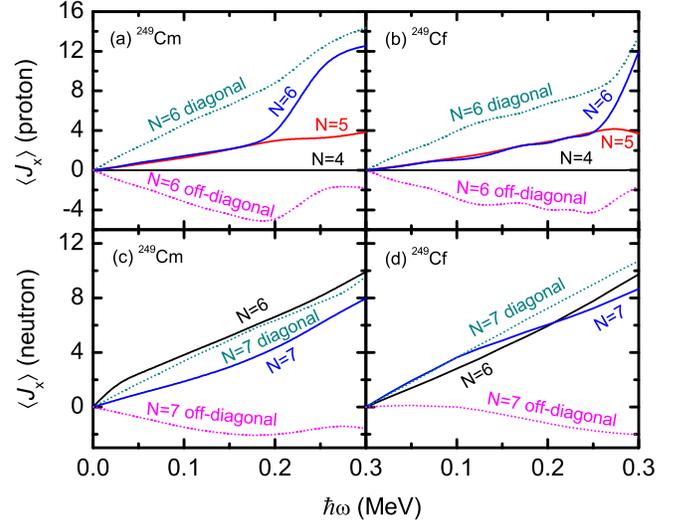}
\caption{\label{fig:jxshell}%
(Color online) Contribution of each proton and neutron major shell to
the angular momentum alignment $\langle J_x\rangle$ for the gsb's in
${}^{249}$Cm and ${}^{249}$Cf. The diagonal $\sum_{\mu} j_x(\mu)$ and
off-diagonal parts $\sum_{\mu<\nu} j_x(\mu\nu)$ in
Eq.~(\protect\ref{eq:jx}) from the proton $N=6$ and neutron $N=7$
shells are shown by dotted lines.%
}
\end{figure}

The contribution of each proton and neutron major shell to the
angular momentum alignment $\langle J_x\rangle$ for the gsb's in
${}^{249}$Cm and ${}^{249}$Cf are shown in Fig.~\ref{fig:jxshell}.
The diagonal $\sum_{\mu} j_x(\mu)$ and off-diagonal parts
$\sum_{\mu<\nu} j_x(\mu\nu)$ in Eq.~(\protect\ref{eq:jx}) from the
proton $N=6$ and the neutron $N=7$ shells are shown by dotted lines.
Note that in this figure, the smoothly increasing part of the
alignment represented by the Harris formula ($\omega J_0 +\omega ^ 3
J_1$) is not subtracted (\textit{cf.} the caption of
Fig.~\ref{fig:MOI}). It can be seen clearly that the upbendings for
the gsb's in ${}^{249}$Cm at $\hbar\omega_c\sim$ 0.20~MeV and in
${}^{249}$Cf at $\hbar\omega_c\sim$ 0.25~MeV mainly come from the
contribution of the proton $N=6$ shell. Furthermore, the upbending
for the gsb in ${}^{249}$Cm is mainly from the off-diagonal part of
the proton $N=6$ shell, while both the diagonal and off-diagonal
parts of the proton $N=6$ shell contribute to the upbending for the
gsb in ${}^{249}$Cf.

\begin{figure}%[h]
\includegraphics[scale=0.51]{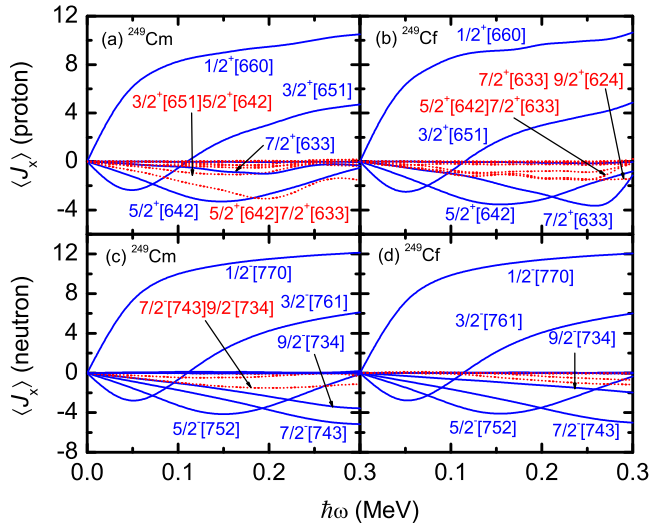}
\caption{\label{fig:jx}%
(Color online) Contribution of each proton orbital in the $N=6$
major shell (top row) and each neutron orbital in the $N=7$
major shell (bottom row) to the angular momentum alignments
$\langle J_x\rangle$ for the gsb's in ${}^{249}$Cm and
${}^{249}$Cf. The diagonal (off-diagonal) part $j_x(\mu)$
[$j_x(\mu\nu)$] in Eq.~(\protect\ref{eq:jx}) is denoted by blue
solid (red dotted) lines.%
}
\end{figure}

In order to have a  more clear understanding of the upbending
mechanism, the contribution of intruder proton orbitals
$i_{13/2}$ (top row) and intruder neutron orbitals $j_{15/2}$
(bottom row) to the angular momentum alignments $\langle
J_x\rangle$ are presented in Fig.~\ref{fig:jx}. The diagonal
(off-diagonal) part $j_x(\mu)$ [$j_x(\mu\nu)$] in
Eq.~(\protect\ref{eq:jx}) is denoted by blue solid (red dotted)
lines. Near the proton Fermi surfaces of Cm and Cf isotopes,
the proton $i_{13/2}$ orbitals are $\pi 3/2^+[651]$, $\pi
5/2^+[642]$ and $\pi 7/2^+[633]$. Other orbitals of $\pi
i_{13/2}$ parentage are either fully occupied or fully empty
(\textit{cf.} Fig.~\ref{fig:Occupation}) and have no
contribution to the upbending. In Fig.~\ref{fig:jx}(a), the PNC
calculation shows that after the upbending ($\hbar \omega \geq$
0.20 MeV) the off-diagonal part $j_x\left(\pi 5/2^+[642] \pi
7/2^+[633]\right)$ changes a lot. The alignment gain after the
upbending mainly comes from this interference term. The
off-diagonal part $j_x\left(\pi 3/2^+[651] \pi
5/2^+[642]\right)$ and the diagonal part $j_x\left(\pi
7/2^+[633]\right)$ also contribute a little to the upbending in
$^{249}$Cm. From Fig.~\ref{fig:jx}(b) one finds that for
$^{249}$Cf the main contribution to the alignment gain after
the upbending comes from the diagonal part $j_x\left(\pi
7/2^+[633]\right)$ and the off-diagonal part $j_x\left(\pi
5/2^+[642] \pi 7/2^+[633]\right)$. Again this tells us that the
upbending in both cases is mainly caused by the $\pi i_{13/2}$
orbitals. The absence of the alignment of $j_{15/2}$ neutrons
in nuclei in this mass region can be understood from the
contribution of the intruder neutron orbitals ($N=7$) to
$\langle J_x \rangle$. For the nuclei with $N\approx150$, among
the neutron orbitals of $j_{15/2}$ parentage, only the
high-$\Omega$ (deformation aligned) $\nu 7/2^-[743]$ and $\nu
9/2^-[734]$ are close to the Fermi surface. The diagonal parts
of these two orbitals contribute no alignment to the upbending,
only the interference terms contribute a little if the neutron
$j_{15/2}$ orbital is not blocked [\textit{cf.}
Fig.~\ref{fig:jx}(c)].

%\section{Summary}

In summary, the recently observed high-spin rotational bands in
odd-$A$ nuclei $^{247, 249}$Cm and $^{249}$Cf~\cite{Tandel2010}
are investigated using the PNC-CSM. In the PNC method for the
pairing correlations, the particle-number is conserved and the
blocking effects are taken into account exactly. The
experimental $\omega$ variations of MOIs and alignments are
reproduced very well by the PNC-CSM calculations. By analyzing
the $\omega$-dependence of the occupation probability of each
cranked Nilsson orbital near the Fermi surface and the
contributions of valence orbitals in each major shell to the
angular momentum alignment, the level crossing and upbending
mechanism in each nucleus is understood clearly. The upbending
in the ground state rotational bands in these nuclei is mainly
caused by the  intruder proton ($N=6$) $\pi i_{13/2}$ orbitals.
The reason of the absence of the alignment of $j_{15/2}$
neutrons is discussed.

%\section{Acknowledgement}

This work has been supported by NSFC (Grant Nos. 10875157 and
10979066), MOST (973 Project 2007CB815000), and CAS (Grant Nos.
KJCX2-EW-N01 and KJCX2-YW-N32). The computation of this work
was supported by Supercomputing Center, CNIC of CAS. Helpful
discussions with G. G. Adamian, N. V. Antonenko, X. T. He, and
F. Sakata are gratefully acknowledged.

%\end{CJK*}


\end{document}